\documentclass[amsmath, amssymb, preprintnumbers, showpacs, showkeys,aps,prb,superscriptaddress,twocolumn]{revtex4-1}

\usepackage{color} 
\usepackage{hyperref}
\usepackage{slashed}
\usepackage{graphicx}
\usepackage{amsmath}
\usepackage{latexsym}
\usepackage{epstopdf}
\usepackage{amsmath}
\usepackage{enumitem}
\usepackage{amssymb,amsmath}
\usepackage{multirow}
\usepackage{mathtools}
\usepackage{bbm}
\usepackage{bm}
\usepackage{amsfonts}
\usepackage{amssymb}
\usepackage{calligra}
\usepackage{calrsfs}
\usepackage{makecell}
\usepackage[normalem]{ulem}
\usepackage[british,UKenglish,USenglish,english,american]{babel}

\begin{document}

\title{Quantum Cluster Quasicrystals}

\author{Guido Pupillo}
\affiliation{University of Strasbourg and CNRS, ISIS (UMR 7006) and IPCMS (UMR 7504), Strasbourg, 67000 France}
\email{pupillo@unistra.fr}

\author{Primo{\v z} Ziherl}
\affiliation{Faculty of Mathematics and Physics, University of Ljubljana, %Jadranska 19, SI-1000
SI-1000 Ljubljana, Slovenia}
\email{primoz.ziherl@ijs.si}
\affiliation{Jo\v{z}ef Stefan Institute, %Jamova 39, SI-1000 
Ljubljana, Slovenia}

\author{Fabio Cinti}
\affiliation{Dipartimento di Fisica e Astronomia, Universit\`a di Firenze, I-50019, Sesto Fiorentino (FI), Italy}
\email{fabio.cinti@unifi.it}
\affiliation{INFN, Sezione di Firenze, I-50019, Sesto Fiorentino (FI), Italy}
\affiliation{Department of Physics, University of Johannesburg, P.O. Box 524, Auckland Park 2006, South Africa}

\begin{abstract}
Quasicrystals remain among the most intriguing materials in physics and chemistry. Their structure results in many unusual properties including anomalously low friction as well as poor electrical and thermal conductivity but it also supports superconductivity, which shows that quantum effects in quasicrystals can be quite unique. We theoretically study superfluidity in a model quantum cluster quasicrystal. Using path-integral Monte Carlo simulations, we explore a 2D ensemble of bosons with the Lifshitz--Petrich--Gaussian pair potential, finding that moderate quantum fluctuations do not destroy the dodecagonal quasicrystalline order. This quasicrystal is characterized by a small yet finite superfluidity, demonstrating that particle clustering combined with the local cogwheel structure can underpin superfluidity even in the almost classical regime. This type of distributed superfluidity may also be expected in certain open crystalline lattices. Large quantum fluctuations are shown to induce transitions to cluster solids, supersolids and superfluids, which we characterize fully quantum-mechanically.
\end{abstract}

%\pacs{pacs here}

\maketitle

%These are the references removed from the abstract \cite{Dubois91}~\cite{Poon92} ~\cite{Kamiya:2018aa}~\cite{Barkan14}.

\section{Introduction}\label{section1}

Despite the order-of-magnitude differences in the characteristic length and energy scales, solid-state and soft matter share many features and phenomena exemplified by the well-established analogy between smectic liquid crystals and superconductors~\cite{deGennes72,Halperin74} and the thriving field of electronic liquid-crystalline mesophases~\cite{Kivelson98,Fernandes:2014aa}. An interesting soft-matter effect with possible analogies at a much smaller scale---e.g., in Rydberg-excited Bose--Einstein condensates~\cite{Pohl2010,Cinti:2014aa}---is clustering. In a good solvent, dilute polymers can be considered as extended finite-size objects but if the concentration is large enough they interpenetrate and overlap. The effective potential between them depends on architecture, functionalization, etc.~\cite{Likos2001} and in some cases, e.g., in amphiphilic dendrimers~\cite{Mladek08}, it may promote ordered phases consisting of evenly spaced multiple-occupancy clusters~\cite{Mladek06,Dotera2014}. Such clustering can be viewed as an instability resulting from a negative component of the Fourier transform of the potential~\cite{Likos01b}.

The physical properties of cluster phases are controlled by their symmetry and by the morphology of the clusters, which may be spherical, cylindrical, sheet-like, inverted, or even bicontinuous~\cite{Shin09}. Both symmetry and morphology are determined by interparticle interaction, and a synthetic approach to generate a host of different phases including cluster quasicrystals (QCs) is based on simultaneous instability at two lengthscales~\cite{Barkan2011,Barkan14,PhysRevB.100.214515}. The cluster phases should all exhibit some degree of activated hopping and the ensuing finite diffusivity~\cite{Moreno07} is expected to be more prominent in the dodecagonal QCs where the neighboring clusters at the perimeter of the characteristic cogwheel-like patches are rather close to each other~\cite{Barkan14}. 

In a quantum system, such dynamics could well lead to novel types of supersolidity, implying the coexistence of (quasi)crystalline and superfluid behavior. To explore this possibility, we theoretically study a 2D ensemble of bosons with the Lifshitz--Petrich--Gaussian pair interaction that produces a classical dodecagonal cluster QC. 

We use path-integral Monte-Carlo (PIMC) simulations to show that the QC remains stable if quantum fluctuations are not too large and that it supports local distributed superfluidity in clusters by a kind of percolating network. In these two respects, it departs from known supersolids~\cite{Pohl2010,PhysRevLett.105.135301,Pfau2016,Santos2016,PhysRevLett.119.215302,PhysRevA.96.013627,Tanzi2019,Chomaz2019}and superglasses~\cite{Biroli08,Yu2012,Carleo2009,Melko2010,Larson2012,Angelone2016}. We show that increased quantum fluctuations induce a series of phase transitions to cluster solids, supersolids and superfluid phases.
%We use path-integral Monte-Carlo (PIMC) simulations to show if, bringing in Bose--Einstein statistic, quantum properties may utterly destroy the  dodecagonal setup or, even more stimulating, to alter the structure by means of multiple phase transitions. Furthermore, if quantum fluctuations are not too large, QC ought to  support local distributed superfluidity in clusters and by a kind of percolating network. In this respects, it departs from known supersolids and superglasses~\cite{Biroli08}.}
Our findings open different possibilities for weak quantum behavior characterized by local superfluidity, say in cluster systems based on honeycomb or Kagome lattices and their 3D analogs. 

%In a quantum system, such dynamics could well lead to novel types of supersolidity~\cite{RevModPhys.84.759}, implying the coexistence of (quasi)crystalline and superfluid behavior. To explore this possibility, we theoretically study a 2D ensemble of bosons with the Lifshitz--Petrich--Gaussian pair interaction that produces a classical dodecagonal cluster QC. We use path-integral Monte-Carlo (PIMC) simulations to show that the QC remains stable if quantum fluctuations are not too large and that it supports local distributed superfluidity in clusters and by a kind of percolating network. In these two respects, it departs from known supersolids and superglasses~\cite{Biroli08}. Our findings open new possibilities for weak quantum behavior characterized by local superfluidity, say in cluster systems based on honeycomb or Kagome lattices and their 3D analogs. 

This article is organized as follows: In Section~\ref{section2} we  introduce the model Hamiltonian, whereas Section~\ref{section3} presents the properties of the dodecagonal QC structure when quantum fluctuations are taken into account. Findings and conclusions are outlined in Section~\ref{section4}.

\section{Model Hamiltonian for quantum cluster quasicrystals}\label{section2}

We consider an ensemble of $N$ two-dimensional bosons of mass $m$ with a many-body Hamiltonian
\begin{equation}\label{eq:ham}
\hat{H}=-\frac{\hbar^2}{2m}\sum_{i=1}^{N}\nabla^2_i+\sum_{i<j}^N V\left(|{\bf r}_i-{\bf r}_j|\right),
\end{equation}
where  
\begin{equation}\label{eq:pot}
V(r) = \exp\left(-\sigma^2r^2/2\right)\sum_{k=0}^{4}C_{2k}\,r^{2k}
\end{equation}
is the Lifshitz--Petrich--Gaussian pair potential~\cite{Barkan14} and ${\bf r}_i \equiv (x_i ,y_i )$ is the position of $i$-th particle. If the parameters $\sigma$ and $C_{2k}$ are chosen such that its Fourier transform features two equal-depth negative minima and the ratio of the corresponding wavevectors is $\sqrt{2+\sqrt{3}}\approx1.93$, this potential leads to a dodecagonal QC pattern in a classical system~\cite{Barkan14}. We too use this particular set of parameters, focusing on quantum effects in the dodecagonal cluster QC. These effects depend on the magnitude of quantum fluctuations encoded by the de Boer parameter~\cite{Sevryuk2010}
\begin{equation}
\Lambda=\sqrt{\frac{\hbar^2}{mr_0^2V_0}},
\end{equation} 
where $V_0=V(0)$ is the pair potential at $r=0$ and $r_0$ is the characteristic length given by the inverse of the wavevector corresponding to the first minimum of the transform of $V(r)$.

\section{12-fold quantum cluster quasicrystal}\label{section3}

\begin{figure}[t!]
\begin{center}
\resizebox{0.99\columnwidth}{!}{\includegraphics{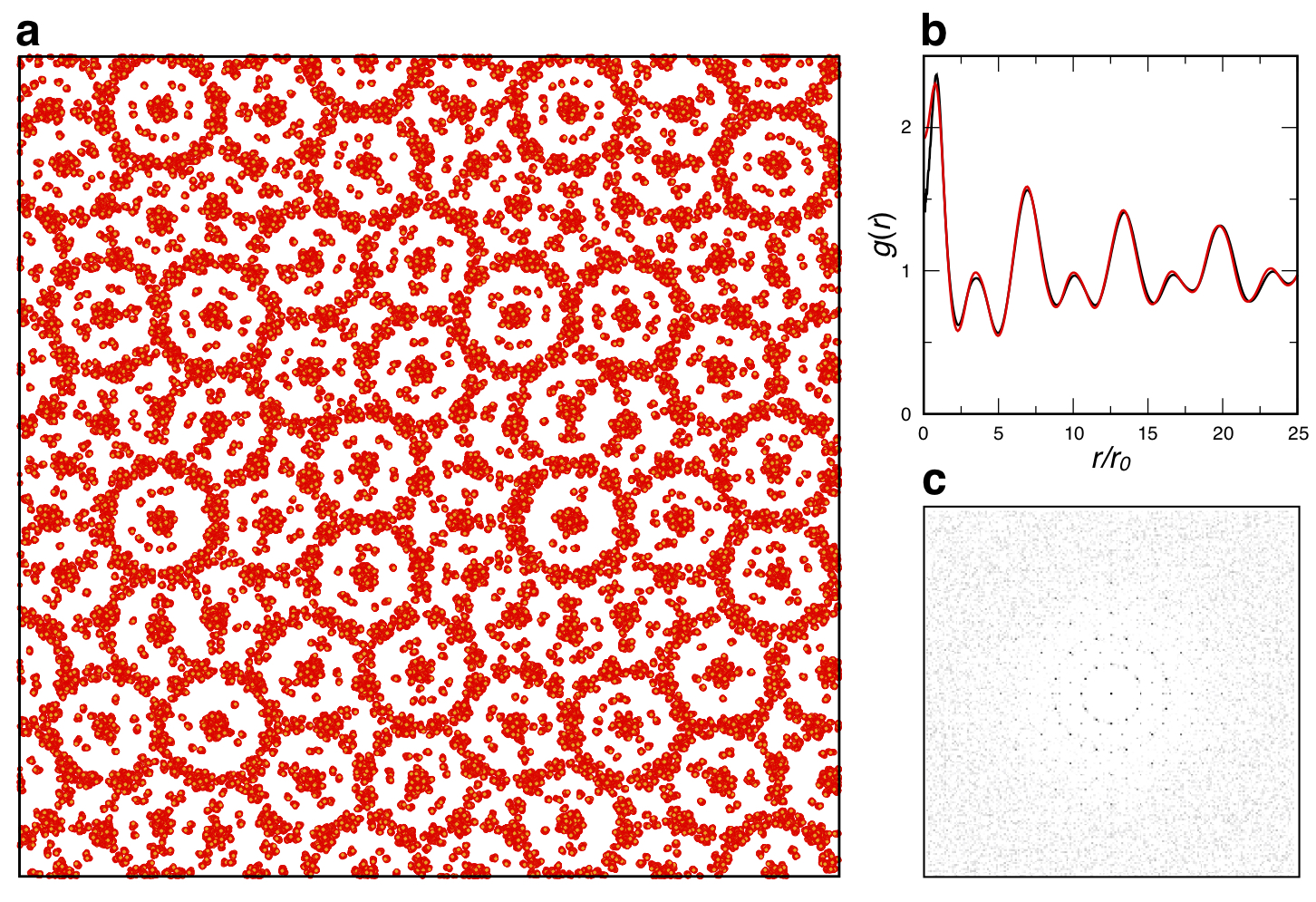}}
\caption{\label{fig1} \textit{Color online}. Quantum QC in the semi-classical regime: {\bf a}, PIMC snapshot of the dodecagonal QC at $\Lambda=0.1$, reduced temperature $t=0.05$, and reduced density $\rho r_0^2=0.8$; here $N = 8192$. {\bf b}, Radial distribution function $g(r)$ for $\Lambda=0.1$ (red line) and $\Lambda=0$ (black line).  {\bf c}, Fourier transform of the positions of world lines in panel a.
The parameters of the interparticle potential in Eq.~\eqref{eq:pot} are 
$\sigma = 0.770746$,  $C_0 = 1$,  $C_2 = -1.09456$,  $C_4 = 0.439744$,  $C_6 = -0.0492739$, and $C_8 = 0.00183183$ as reported in Ref.~\cite{Barkan14}.}
\end{center}
\end{figure}

We employ PIMC simulations~\cite{Ceperley1995} based on the continuous-space worm algorithm~\cite{PhysRevLett.96.070601} to find the equilibrium state of Eq.~(\ref{eq:ham}) at a fixed temperature and a fixed number of particles $N$ (canonical ensemble), with $N$ between 2048 and 8192. All simulations were carried out using periodic boundary conditions along both directions. 
In particular, we study the ensemble at temperatures around the range where the classical dodecagonal QC is stable, first at small $\Lambda$. Figure~\ref{fig1} shows the quantum dodecagonal QC for $\Lambda=0.1$, reduced temperature $t=k_BT/V_0=0.05$, reduced density $\rho r_0^2=0.8$, and $N = 8192$. 
In this figure we focus on the semi-classical limit, i.e., on boltzmannons, where the zero-point motion due to quantum fluctuations is accounted for, whereas the world-line exchanges leading to superfluidity are initially excluded. Panel a shows a snapshot of the projection of world lines onto the $xy$-plane obtained by tracing over the imaginary time evolution; this is a good representation of the square of the semi-classical many-body wave function~\cite{Ceperley1995}. In Fig.~\ref{fig1}a, the paths are essentially localized around the energy minima of the QC structure observed in the classical limit in Ref.~\cite{Barkan14} despite a somewhat larger reduced temperature (0.05 vs.~0.03) and despite quantum fluctuations. 

The similarity of the semi-classical and the classical QCs is further corroborated by the radial distribution functions, $g(r)$ \cite{Allen2017},
in Fig.~\ref{fig1}b which are virtually identical except close to $r=0$: The quantum $g(r)$ is somewhat larger than the classical one, indicating increased local fluctuations of the particles (inset to Fig.~\ref{fig2}e). We note that the introduction of Bose--Einstein statistics further enhances this effect (see below). 

The Fourier transform of the $\Lambda=0.1,t=0.05$ semi-classical dodecagonal QC (Fig.~\ref{fig1}c) evidently has a 12-fold rotational symmetry. Fig.~\ref{fig1}c was obtained by taking the averaged position of each world-line (centroid coordinates) in space. In terms of position, the peaks agree with the stronger inner peaks characterizing the classical counterpart of our QC~\cite{Barkan14} as well as with those seen experimentally in, e.g., a dendrimer-micelle QC~\cite{Zeng04}. The variations in intensity --and especially the presence of the diffuse outer ring-- reflect the different form factor and thus a different intra-cluster structure, as also observed in $g(r)$ at small $r$. 

\begin{figure*}[t!]
\begin{center}
\resizebox{0.8\textwidth}{!}{\includegraphics{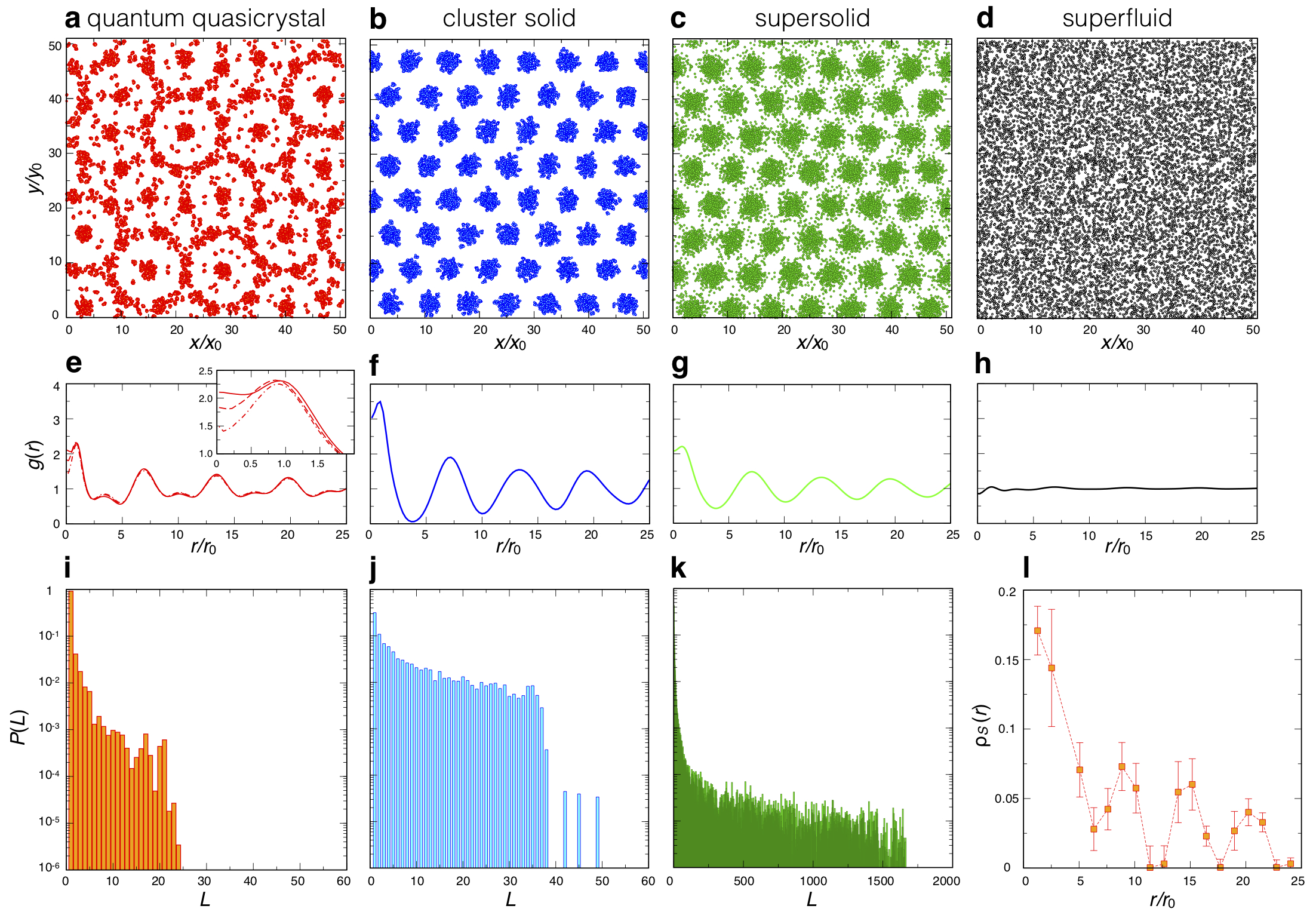}}
\caption{\label{fig2}\textit{Color online}. Quantum quasicrystal and reentrant superfluidity: {\bf a-d}, Dodecagonal quantum QC, cluster crystal, cluster supersolid, and superfluid at $t=0.05$,  $\rho r_0^2=0.8$, and $\Lambda=0.1,0.141,0.632,$ and 1, respectively; $N=2048$. {\bf e-h}, Radial distribution functions of the four phases in panels a-d. Inset in {\bf e}, distribution functions of QC for bosons (solid line), semiclassical boltzmannons (dashed line) and classical particles (dot-dashed line). {\bf i-k}, Frequency of exchange cycles of length $L$ in the QC, cluster solid, and supersolid in panels a-c. {\bf l}, Superfluid density profile of the dodecagonal QC from panel a.}
\end{center}
\end{figure*}

Full quantum effects combining fluctuations and bosonic statistics are investigated in Fig.~\ref{fig2}. As the de Boer parameter is increased at $t=0.05$ and $\rho r_0^2=0.8$, the ensemble undergoes three transitions (Fig.~\ref{fig2}a-d). At $\Lambda\approx0.12$, the dodecagonal QC shown in panel a is replaced by a hexagonal cluster crystal (panel b); the clusters are well-defined and evidently larger than those in the QC, their spacing being the same as the radius of the dodecagonal wheels in the QC. Given that the pair potential features two local minima~\cite{Barkan14}, the increase of cluster size and their rearrangement suggest that at the larger $\Lambda$, intra-cluster quantum fluctuations render the smaller-distance minimum less effective. The structural differences between the two phases readily show in the radial distribution functions (Fig.~\ref{fig2}e and f). The modulation of $g(r)$ in the cluster solid is very prominent, virtually vanishing between nearest-neighbor clusters, whereas in the QC it is considerably smaller.

We now turn to the quantum properties of the QC and the cluster solid, first monitoring  
the frequency of cycles of permutations involving $L$ bosons denoted by $P(L)$, with $1\leq L\leq N$. 
The occurrence of long permutation cycles in the histogram $P(L)$ should be linked to the existence of finite superfluid response throughout the system. In the $\Lambda=0.1$ QC, the distribution of $P(L)$ stretches to $L\approx25$ (Fig.~\ref{fig2}i). Since the clusters contain about 18 particles, this implies a finite particle exchange between clusters, pointing to a local distributed superfluidity in this {\sl quantum} QC phase. In the cluster solid (Fig.~\ref{fig2}j), permutation cycles stay within single clusters, which contain about 36 particles.

The competition between the tendency of bosons to delocalize at low temperatures and the structure of QC and cluster crystal is also reflected in the superfluid density. In the PIMC framework, the superfluid density $\rho_s$ is evaluated by applying the linear response theory to address the response of the boundary motion of the ensemble. 
In this study, superfluidity was sampled by using the expression $\rho_s=(4\rho m^2)/(\hbar^2\beta I_{cl})\left\langle A^2\right\rangle$, where $\beta=1/k_BT$ and $I_{cl}$ is the classical moment of inertia of the particles calculated with respect to the axis perpendicular to the $xy$-plane. In the context of the path-integral formalism, the estimator $A$ then gives the world-line area of closed particle trajectories projected onto the $xy$ plane. 
 Likewise, the local contribution to the superfluid density, $\rho_s(r)$ \cite{kwon2006,Jain2011}, is obtained by sampling the radial dependence of the local area operator $ A\cdot A(r)$ and the corresponding local moment of inertia $I_{cl}(r)$. In a true periodic structure (see, for instance, the supersolid phase in Fig.~\ref{fig2}c) the evaluation of $\rho_s$ using the area estimator techniques gives results that are fully consistent with the well-known ``winding number" estimator\cite{poll87}.  
%The average superfluid density and the local averaged superfluid density $\rho_s(r)$ are computed using the area estimator procedure~\cite{Ceperley1995} (see Methods). 

In the QC in Fig.~\ref{fig2}a, $\rho_s$ is small but finite, the fraction of superfluid particles being about 1$-$2\% consistent with the measured exchange cycles and a picture of weak distributed superfluidity. On the other hand, the global superfluid signal is completely suppressed in both QC and cluster crystal.

Figure~\ref{fig2}l shows the superfluid density profile in the QC, which is evidently nonuniform. By comparing the profile with the snapshot and $g(r)$ in Fig.~\ref{fig2}a and e, respectively, we see that $\rho_s(r)$ is small but non-negligible both in the central clusters of the dodecagonal wheels and around their perimeter. In fact, the local superfluid signal is non-zero in accordance with the structure of Fig.~\ref{fig2}a. Consistent with quantum-mechanical exchanges shown in Fig.~\ref{fig2}j, in the cluster crystal $\rho_s(r)$ should be finite (about unity as $t\to0$) inside each cluster and zero otherwise. Given the size of our system, these results are not affected by finite-size effects. We note that a full finite size scaling of $\rho _s$ is here not possible, as increasing the already very large $N$ makes computations exceedingly long at sufficiently low $T$, while decreasing $N$ changes the QC structure in favour of a $\sigma $-type one. We find comparable superfluid signal in this latter case.

At even larger $\Lambda$, fluctuations become even more prominent and at a $\Lambda\approx0.54$ they lead to the transition from the cluster solid to the supersolid. In the latter, the diameter of the clusters is larger than in the cluster solid whereas the lattice spacing remains the same. This facilitates delocalization and long exchanges of particles hinted at by the many particles seen between the clusters in the $\Lambda=0.632$ snapshot in Fig.~\ref{fig2}c and proven by the distribution of the exchange cycles in Fig.~\ref{fig2}k, which includes cycles with over 1500 bosons in an ensemble of $N=2048$ particles. The superfluid fraction of the supersolid in Fig.~\ref{fig2}c is less than unity, amounting to $\rho_s\approx0.46$, as expected for a spatially modulated superfluid~\cite{Leggett1970,PhysRevLett.105.135301,Henkel2012,PhysRevA.87.061602}, and remains almost unchanged upon cooling. Particle delocalization is further seen in $g(r)$ in Fig.~\ref{fig2}g where the maxima and minima are at the same positions as in the cluster solid but are much less prominent. At $\Lambda\approx0.8$, the strong world line delocalization turns the ensemble into a homogeneous superfluid where the superfluid fraction is 1. Figure~\ref{fig2}d shows a snapshot of this phase at $\Lambda=1$, and the corresponding radial distribution function in Fig.~\ref{fig2}h is almost featureless.

The phase diagram of the $N=2048$ ensemble at a density of $\rho r_0^2=0.8$ is shown in Fig.~\ref{fig3}. The large-$\Lambda$ region is occupied by the superfluid whereas the large-$t$ region belongs to the normal fluid. The region stretching roughly to $t=0.1$ and $\Lambda=0.7$ is divided among three solid phases, with the normal dodecagonal QC present only at vanishing $\Lambda$s. The quantum QC/cluster solid/supersolid/superfluid sequence is representative of $t$s between about $0.03$ and 0.1, whereas at $t$s below 0.03 the quantum QC phase is absent. Interestingly, at low $t$ and for $\Lambda \lesssim 0.2$, quantum fluctuations do not stabilize a QC but rather strengthen the occurrence of the cluster crystal in agreement with the classical case\cite{Barkan14}.

\section{Conclusions}\label{section4}

\begin{figure}[t]
\begin{center}
\resizebox{0.98\columnwidth}{!}{\includegraphics{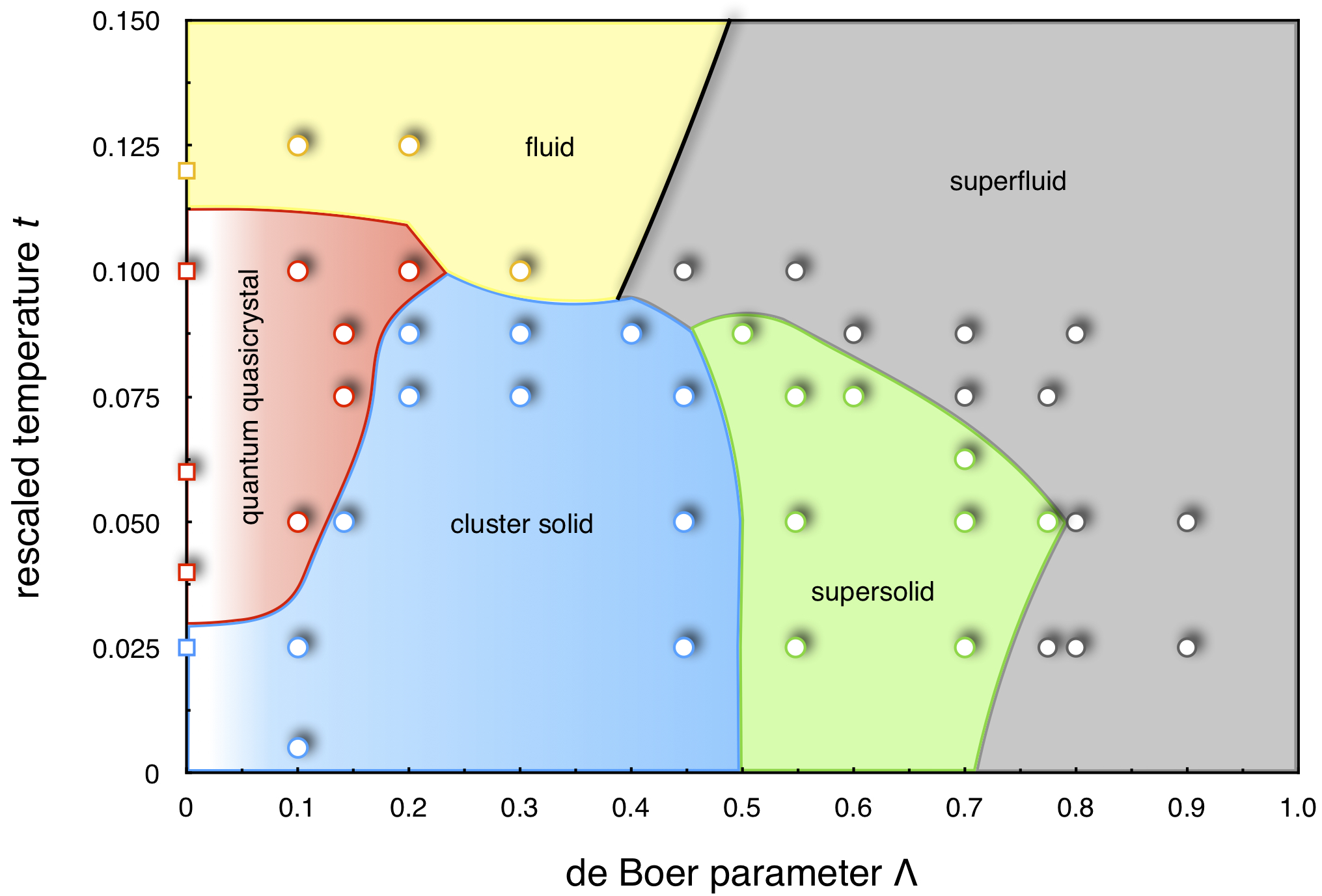}}
\caption{\label{fig3}\textit{Color online}. Phase diagram of 2D Lifshitz--Petrich--Gaussian bosons featuring the dodecagonal quantum QC, supersolid, and superfluid as well as a hexagonal cluster solid and a normal fluid phase. Circles represent the $N=2048$ datapoints analyzed; the indicated phase boundaries are approximate. The thick black line depicts the Berezinskii--Kosterlitz--Thouless transition $t_{BKT}\approx\pi \bar{\rho}_sr_0^2\Lambda^2/2$ ($\bar{\rho}_s$ being the superfluid density at $t_{BKT}$) between the normal fluid (yellow) and the superfluid phase (gray).}
\end{center}
\end{figure}

This phase diagram shows the reentrant nature of superfluidity in our system. With a proper structural support---here the dodecagonal QC---superfluid behavior can be extended down to small values of the de Boer parameter, albeit in a fraction of particles rather than globally like in the large-$\Lambda$ superfluid phase. There may exist other non-close-packed 2D lattices that could host distributed superfluidity, say a honeycomb or Kagome lattice with vertex figures 6.6.6 and $(3.6)^2$, respectively. Figure~\ref{fig3} is also important because it provides an additional insight into the mechanism of structure formation in pair potentials with equal-depth-minima transforms, showing that the temperature range where the desired structure is stable is reasonably broad but that at large and very small $t$s it is replaced by the fluid and the energy-minimizing phase, respectively. Finally, our phase diagram was computed at a fixed particle density. 
Given that in the classical repulsive coreless cluster-forming systems the phase sequence is qualitatively the same at all densities\cite{Mladek06,Glaser07}, we expect that at a somewhat larger or smaller density our phase diagram too is simply rescaled but otherwise unaltered.

Our quantum QC is a novel self-assembled phase with local and distributed superfluidity close to the classical regime, which contributes to the advances at the crossroad between quasiperiodic systems and quantum phenomena illustrated by, e.g., topological states in quasicrystals\cite{Kraus2012}, the Dirac electrons in dodecagonal graphene\cite{Ahn2018}, and time quasicrystals\cite{Autti2018}. Our results emphasize that this complex behavior can result solely from pair interactions, and it would be interesting to search for it in other classes of two-lengthscale soft-core potentials\cite{Barkan2011,Archer2013} as well as in experiments. The recent observation of self-assembled supersolid behavior in 1D with dipolar magnetic atoms\cite{Bottcher2019,Tanzi2019,Chomaz2019}, similar to that predicted in cluster-forming interactions, raises the question of whether self-assembled QC behavior may be engineered in such systems---possibly aided by structured optical potentials such as those used in cold-atom experiments where superfluidity is furnished by a Bose--Einstein condensate trapped in a laser-generated lattice\cite{PhysRevLett.111.185304,Viebahn2019}, 
or even condensate featuring a Rashba spin-orbit coupling combined with dipolar interactions \cite{PhysRevLett.120.060407}.

%{\bf Methods} In relation to the latter, we evaluated its global as well as local estimator $\rho_s$ and $\rho_s(r)$, respectively\cite{kwon2006,Sindzingre1989,Jain2011}. In quasi-periodic structures such as the QC phase discussed here, superfluidity was sampled by using the expression $\rho_s=(4\rho m^2)/(\hbar^2\beta I_{cl})\left\langle A^2\right\rangle$, where $\beta=1/k_BT$ and $I_{cl}$ is the classical moment of inertia of the particles calculated with respect to the axis perpendicular to the $xy$-plane. In the context of the path-integral formalism, the estimator $A$ then gives the world-line area of closed particle trajectories projected onto the $xy$ plane. Likewise, the local contribution to the superfluid density is obtained by sampling the radial dependence of the local area operator $ A\cdot A(r)$ and the corresponding local moment of inertia $I_{cl}(r)$. In a true periodic structure (see, for instance, the supersolid phase in Fig.~\ref{fig2}c) the evaluation of $\rho_s$ using the area estimator techniques gives results that are fully consistent with the well-known ``winding number" estimator\cite{poll87}.  

\textit{Acknowledgements.} %\noindent
We thank T.~Dotera, M.~Engel, R.~Lifshitz, T.~Macr\`i, T.~Pohl and S.~Pilati for valuable discussions. 
The authors acknowledge the financial support from the Slovenian Research Agency (research core funding No. P1-0055), the French ANR - ERA-NET QuantERA - Projet RouTe (ANR-18-QUAN-0005-01), Labex NIE, IUF and USIAS. 

%%%%%%%%%%%%%%%%%%%%%%%%%%%%%%%%%%%%%%%
%%%%%%%%%%%%%%%%%%%%%%%%%%%%%%%%%%%%%%%
%%%%%%%%%%%%%%%%%%%%%%%%%%%%%%%%%%%%%%%
%\bibliographystyle{apsrev4-1} 
%\bibliography{bose}

%merlin.mbs apsrev4-1.bst 2010-07-25 4.21a (PWD, AO, DPC) hacked
%Control: key (0)
%Control: author (72) initials jnrlst
%Control: editor formatted (1) identically to author
%Control: production of article title (-1) disabled
%Control: page (0) single
%Control: year (1) truncated
%Control: production of eprint (0) enabled
%

\end{document}